\documentclass{ceab}   

\usepackage{subcaption}
\usepackage{float}
\usepackage[font=footnotesize]{caption}

\usepackage{epsfig}     
\usepackage{graphicx}   
\usepackage{url}
\usepackage{natbib}     
\usepackage[T1]{fontenc} 
\usepackage{babel}       
    
\def\runningtitle{R. Camphyn ECRS2024 Proceedings}   
\def\articlenumber{7}
\def\ppages{1--4}

\begin{document}

\title{Offline Neutrino Filtering using a Convolutional Neural Network-Based
Algorithm at the Radio Neutrino Observatory Greenland}

\author{Camphyn R.$^{1,2}$ for the RNO-G Collaboration
\vspace{2mm}\\
\it $^1$ Université Libre de Bruxelles - Inter-University Institute For High Energies\\
\it $^2$Fonds de la Recherche Scientifique
}

\maketitle

\begin{abstract}
Neutrino astronomy is a vibrant field of study in astrophysics, offering unique insights into the Universe's most energetic phenomena. The combination of a low cross section and zero electromagnetic charge ensure
that a neutrino retains most information about its original source while traversing the universe. On
the other hand, these low cross sections, combined with a reduced flux at higher energies, make the
neutrino one of the most elusive particles to detect in the standard model. The Radio Neutrino
Observatory in Greenland (RNO-G) aims to detect sporadic neutrino interactions in the Greenlandic ice
sheet by means of electromagnetic signals in the radio frequency range, induced by the produced charged
secondary particles. The low incoming neutrino flux forces the detector to set a low trigger threshold,
leading to the measured data being overwhelmed by thermal noise fluctuations. Hence, a
sophisticated and robust filter is needed to differentiate between neutrino-like signals and noise. In
this contribution we present the current work on the development of such a filter, based on a convolutional neural network architecture. The network employed uses real RNO-G
data and simulated neutrino signals to categorize measured data as noise or neutrino-like events.
\end{abstract}


\section{Introduction}

Some of the most pressing questions in astroparticle physics relate to the sources and production mechanisms behind ultra-high-energy cosmic rays (UHECRs). These charged particles reach energies up to 100 EeV (= $10^{20}$ eV) and can interact with the matter and radiation surrounding their source, yielding interaction products in the form of photons and neutrinos. This implies that UHECR sources are visible through three distinct messengers, if one includes the cosmic rays that are able to escape the source environment. The neutrino has a very low probability to interact with the medium through which it travels. An astrophysical neutrino detection contains information on a source's location in the sky and the physics through which a source produces and accelerates UHECRs.

The first detector to observe an astrophysical neutrino flux was the IceCube Neutrino Observatory \citep{Icecube1, Icecube2}. IceCube instruments one cubic kilometer of Antarctic ice to detect the faint Cherenkov light produced by neutrino-induced secondary leptons as they traverse the detector. The Radio Neutrino Observatory Greenland (RNO-G) \citep{RNO-G:2020rmc}, currently under construction, uses radio waves produced by an incoming neutrino-induced particle cascade through the Askaryan effect \citep{Askaryan:1961pfb}. When completed and fully operational, this novel detector will work with electromagnetic, gravitational wave and cosmic ray observatories to help construct a full theory on the sources and mechanisms behind UHECRs, exploiting the power of multimessenger astronomy.

\section{The Radio Neutrino Observatory in Greenland}

\begin{figure}[h!]
    \centering
    \includegraphics[scale = 1.]{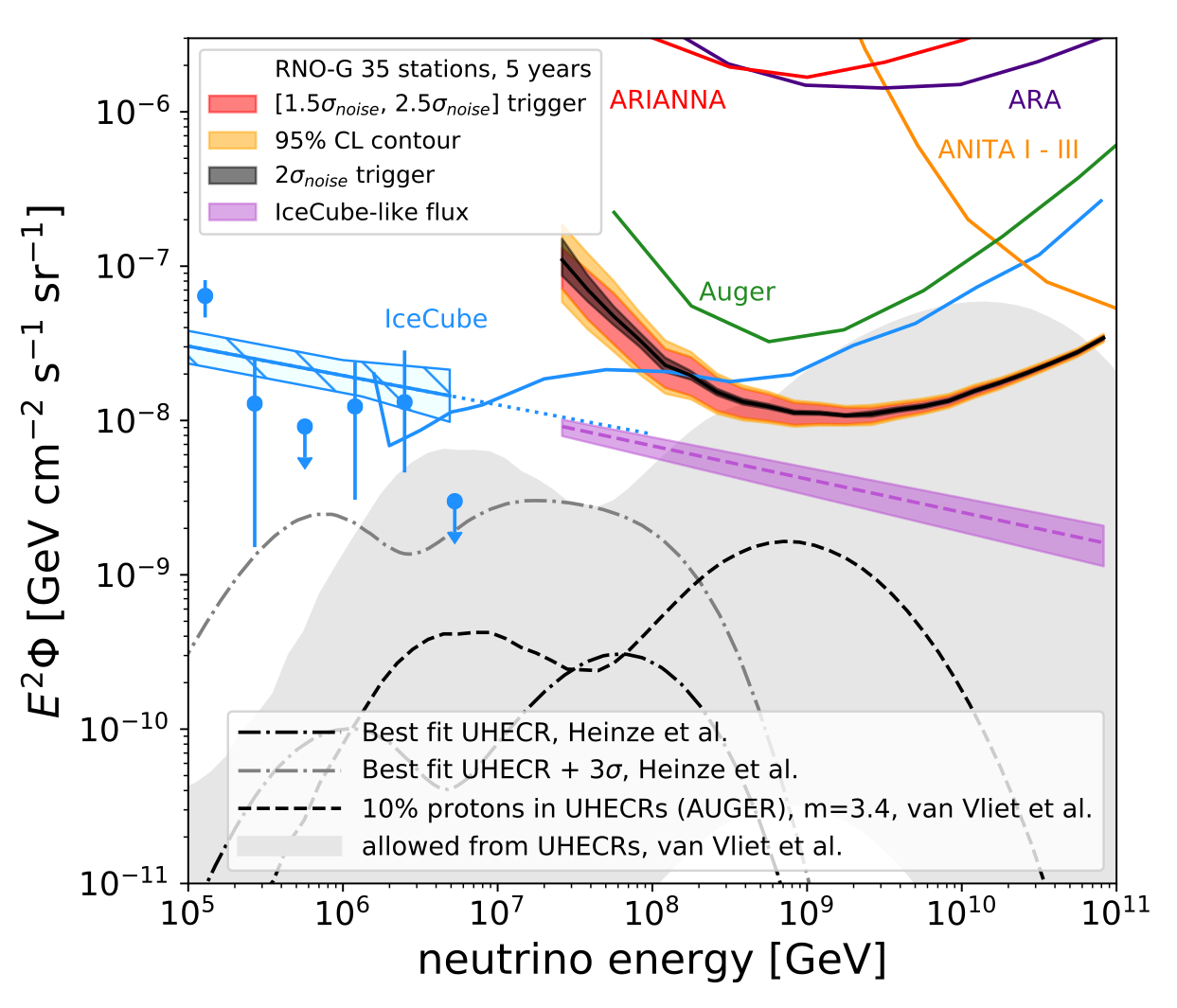}
    \caption{Predicted sensitivity of RNO-G compared to neutrino flux models (gray areas and black-dashed lines) and upper limits set by previous experiments (colored solid lines). The IceCube-like flux in purple and blue-dashed is the predicted neutrino flux in the case that the flux at higher energies is an extension of the fluxes observed at lower energies. Figure taken from \cite{RNO-G:2020rmc}}
    \label{fig:RNOG_sens}
\end{figure}

RNO-G will be the first large scale in-ice radio neutrino detector, enabling the potential discovery of the first ultra-high-energy neutrino. This would extend the known neutrino spectrum beyond IceCube's current measurements. The predicted sensitivity curve is illustrated in Figure \ref{fig:RNOG_sens}.

\begin{figure}[h!]
    \centering
    \includegraphics[width=\linewidth]{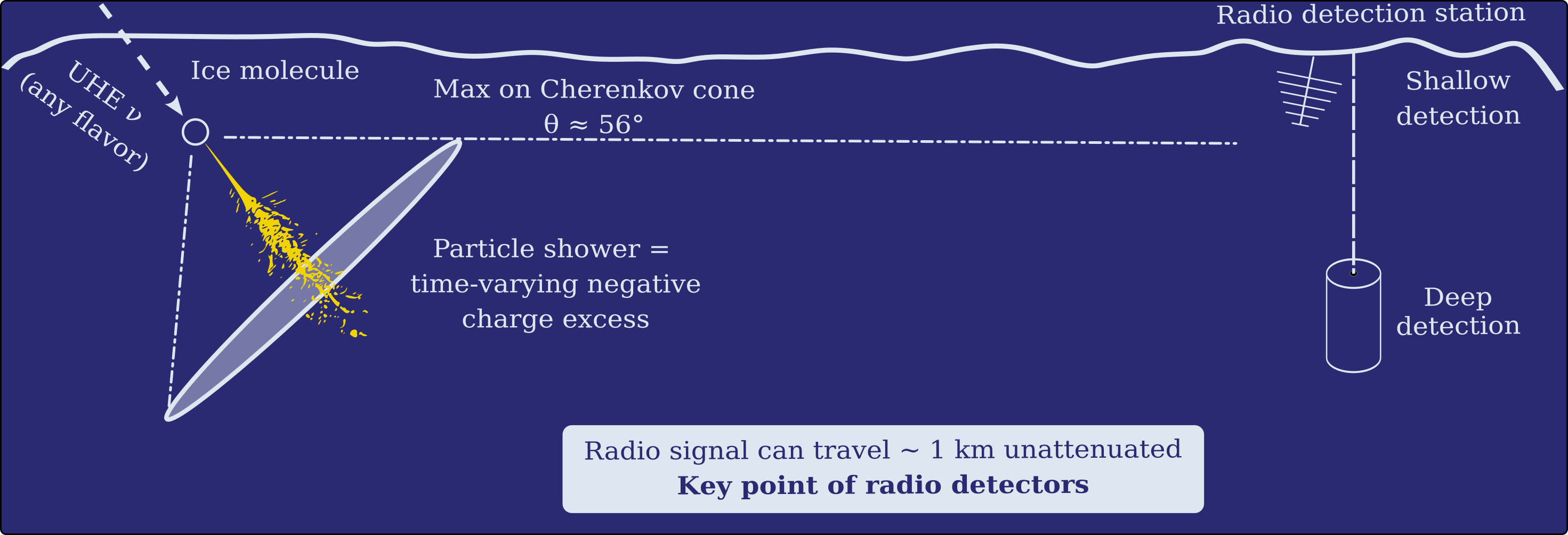}
    \caption{Simplified illustration of a high energy neutrino producing electromagnetic radiation in the radio frequency range through the Askaryan effect.}
    \label{fig:askaryan}
\end{figure}

RNO-G will operate by measuring radio waves emitted by a neutrino-induced particle cascade in the Greenlandic ice sheet. The process yielding these radio waves is dubbed the Askaryan effect. When a neutrino enters the Greenlandic ice sheet, it can interact with an ice molecule and start a particle cascade. As this cascade propagates through the ice, it develops a charge asymmetry, owing to both positron annihilation and the ionization of the traversed ice volume. This moving negative charge emits radiation in the radio frequency range. The propagation speed of the cascade exceeds the speed of light in ice, giving rise to a Cherenkov cone, on which the emitted radio waves interact constructively. A cartoon representation of this process is shown in Figure \ref{fig:askaryan}.

\begin{figure}[h!]
\centering
\begin{subfigure}{0.5\textwidth}
    \centering
    \includegraphics[width=0.9\textwidth]{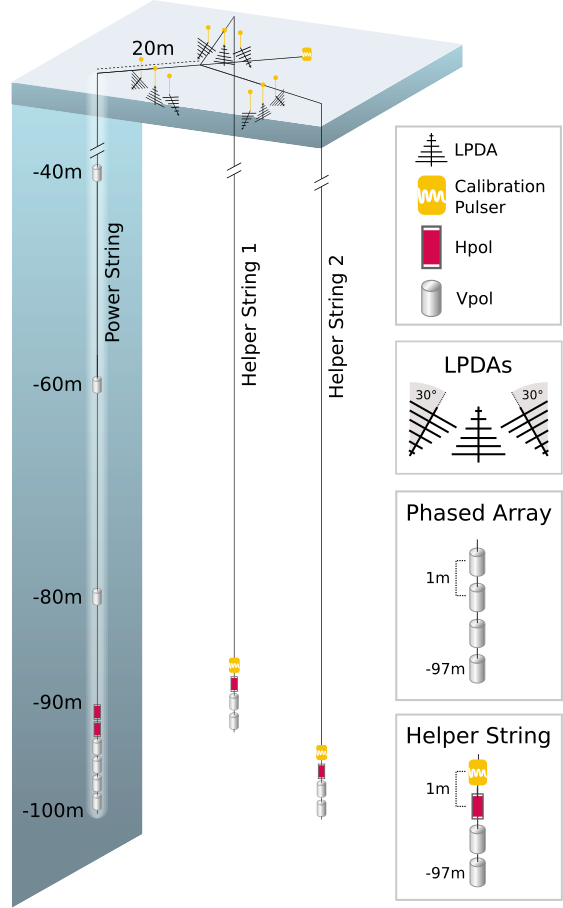}
    \caption{Schematic of one RNO-G station.}
\end{subfigure}%
\begin{subfigure}{0.5\textwidth}
    \centering
    \includegraphics[scale=0.3]{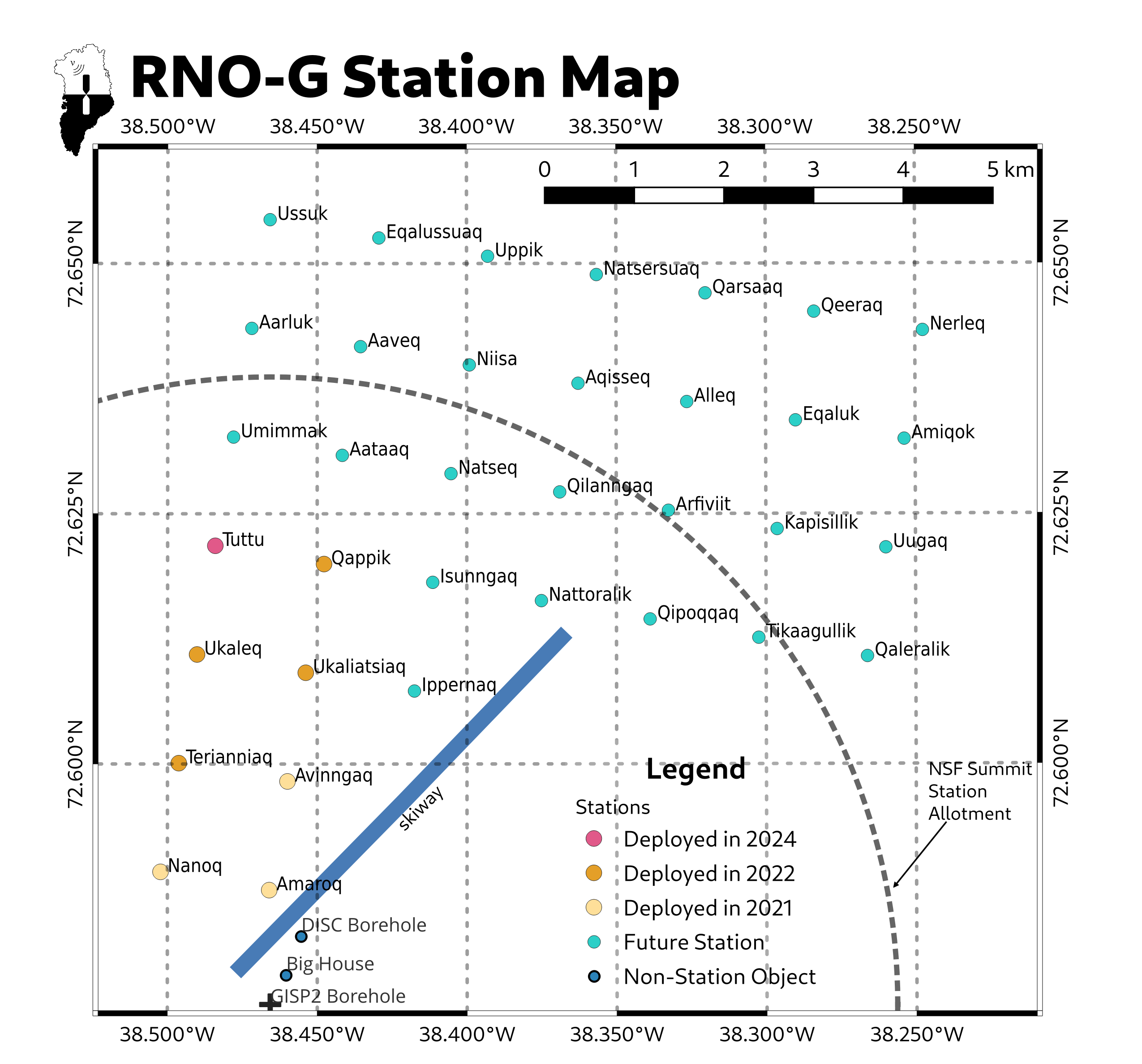}
    \caption{The full RNO-G station array.}
\end{subfigure}
\caption{Figure adopted from \cite{RNOGfirstseven}}
\label{fig:station}
\end{figure}

RNO-G aims to detect these radio waves by instrumenting an area of around $50$ km$^{2}$ by drilling $\sim100$ m into the ice and deploying radio antennas. The full RNO-G layout will consist of 35 independent stations, each containing 15 in-ice antennas and 9 surface antennas. This hybrid design is informed by two predecessor experiments, ARIANNA \citep{ARIANNA:2014fsk} and ARA \citep{ARA} and can be seen in Figure \ref{fig:station}. Every station is powered by solar panels, but the addition of wind turbines would extend the detector's up-time through the Greenlandic winter. These wind turbines need to be manufactured specifically to withstand the hostile polar dessert conditions. Two stations already support wind turbines and more are planned to be added.

The string designated as "power string" contains the physics trigger in the form of a phased array. The phased array consist of 4 vertically polarized antennas on which an artificial time delay can be applied to perform an interferometric plane wave search.

The location of RNO-G at Summit Station Greenland serves a dual purpose. On the high-energy end of the neutrino spectrum, RNO-G's field of view will complement that of IceCube. RNO-G can detect high-energy neutrinos in the Northern Hemisphere. To these neutrinos, the Earth would appear too opaque to reach IceCube's location on the South Pole. Lower energy neutrino detections made by IceCube could be enhanced by RNO-G when investigating whether a higher energy component was blocked from IceCube's field of view by the Earth.

Currently, 8 stations have already been deployed and are taking science data. The construction, commissioning and operation of RNO-G will inform the design of the next generation of neutrino observatories, such as IceCube-Gen2 \citep{IceCube-Gen2:2020qha}.

\section{Neural network based event filtering}

The current trigger rate of RNO-G is $\sim$1 Hz. Out of these triggers, only a handful over a period of 5 years are expected to contain neutrino events. The remainder are mostly thermal noise events, caused by the blackbody radiation from the ice as well as the detector electronics. Since the trigger rate is so high, the full dataset of measured events needs to be reduced to a manageable size before applying sophisticated event reconstruction algorithms. This proceedings describes an offline filter that can perform this reduction by training a convolutional neural network to correctly identify and tag thermal noise events. Note that the designation "noise" in this context does not include physical processes that mimic a neutrino signal, such as cosmic rays. 

Since this is a supervised binary classification problem, a dataset with appropriate labels is to be used. The two labels are ``neutrino signal with noise'' and ``purely noise'', respectively corresponding to the integers 1 and 0. The network is constructed to assign an event a score ranging between $0$ and $1$. An event is defined as one full readout\footnote{With a sampling rate of $3.2$ GHz over $2048$ samples, this comes down to a time window of $640$ ns} of all 24 antennas of one RNO-G station.

\begin{figure}[h!]
\centering
\begin{subfigure}{.5\textwidth}
  \centering
  \includegraphics[width=\linewidth]{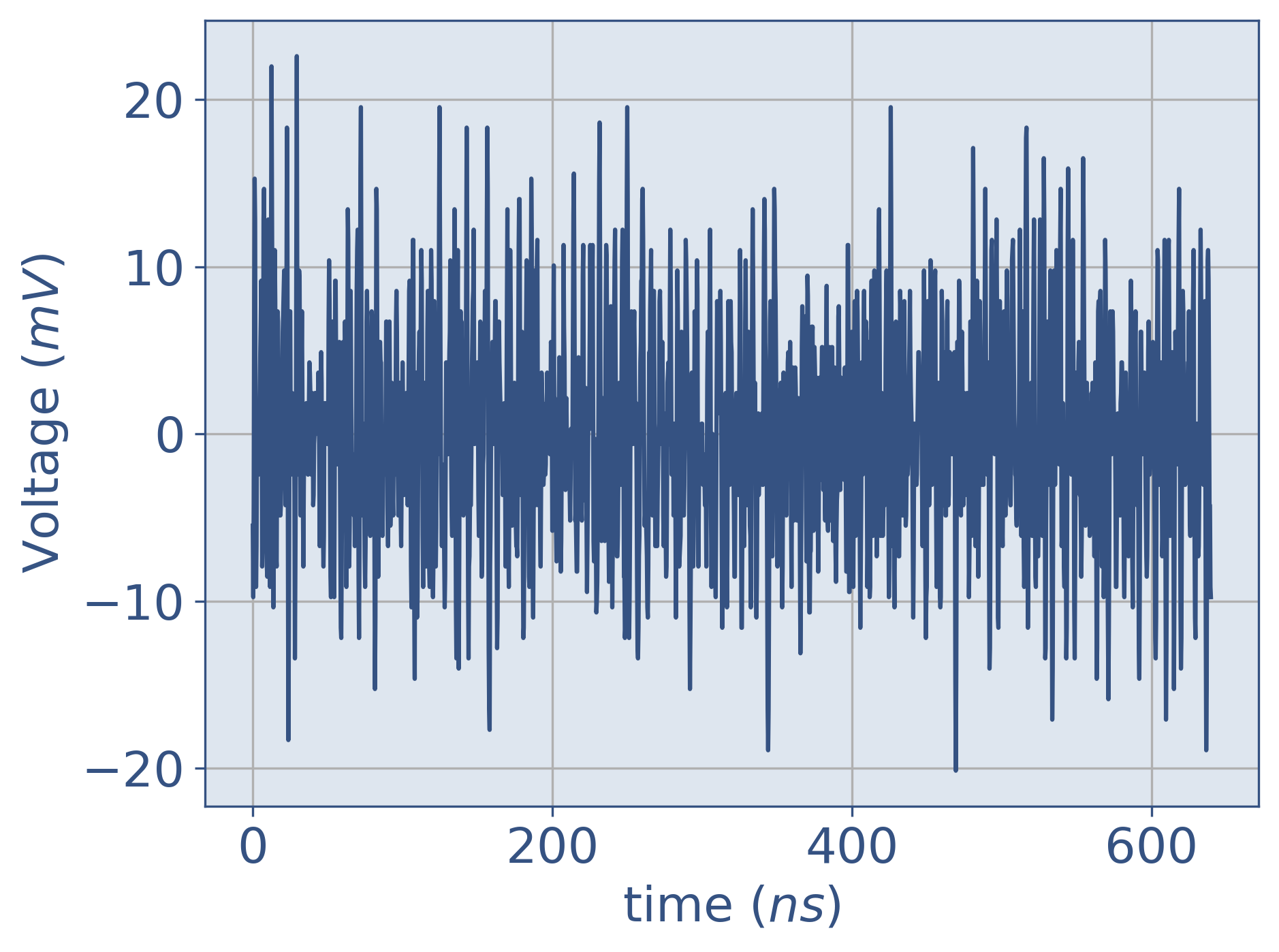}
  \caption{Time trace from a noise event.}
  \label{fig:exnoise}
\end{subfigure}%
\begin{subfigure}{.5\textwidth}
  \centering
  \includegraphics[width=\linewidth]{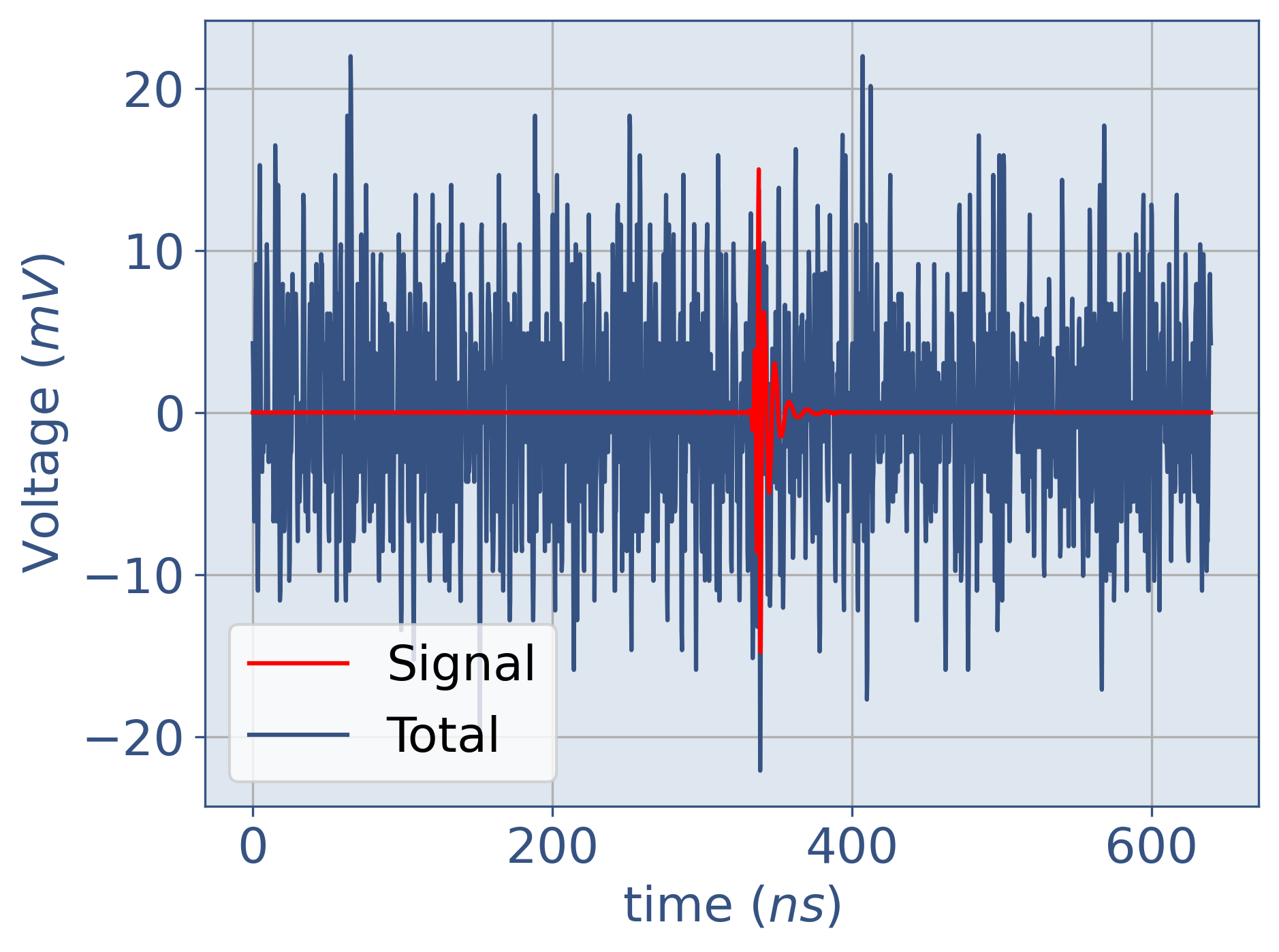}
  \caption{Time trace from a noisy signal event.}
  \label{fig:exsignal}
\end{subfigure}
\caption{Illustrative example of time traces taken from (a) a purely noise event and (b) a signal convoluted with noise, as would be seen in a single vertically polarized antenna in one RNO-G station.}
\label{fig:noise_signal}
\end{figure}

The noise label was populated with data triggered on thermal noise fluctuations. The signal label requires the use of simulated neutrino signals. To obtain a neutrino signal mixed with noise, pure neutrino signal simulations were superimposed on software-triggered\footnote{At regular intervals, forced triggers are recorded to help characterize the noise environment.} events. The simulations were made using the open-source simulation software package NuRadioMC \citep{Glaser:2019cws}. Figures \ref{fig:exnoise} and \ref{fig:exsignal} present two example time traces, each corresponding to one of the two classes.

\begin{figure}[h!]
    \centering
    \includegraphics[width=0.6\linewidth]{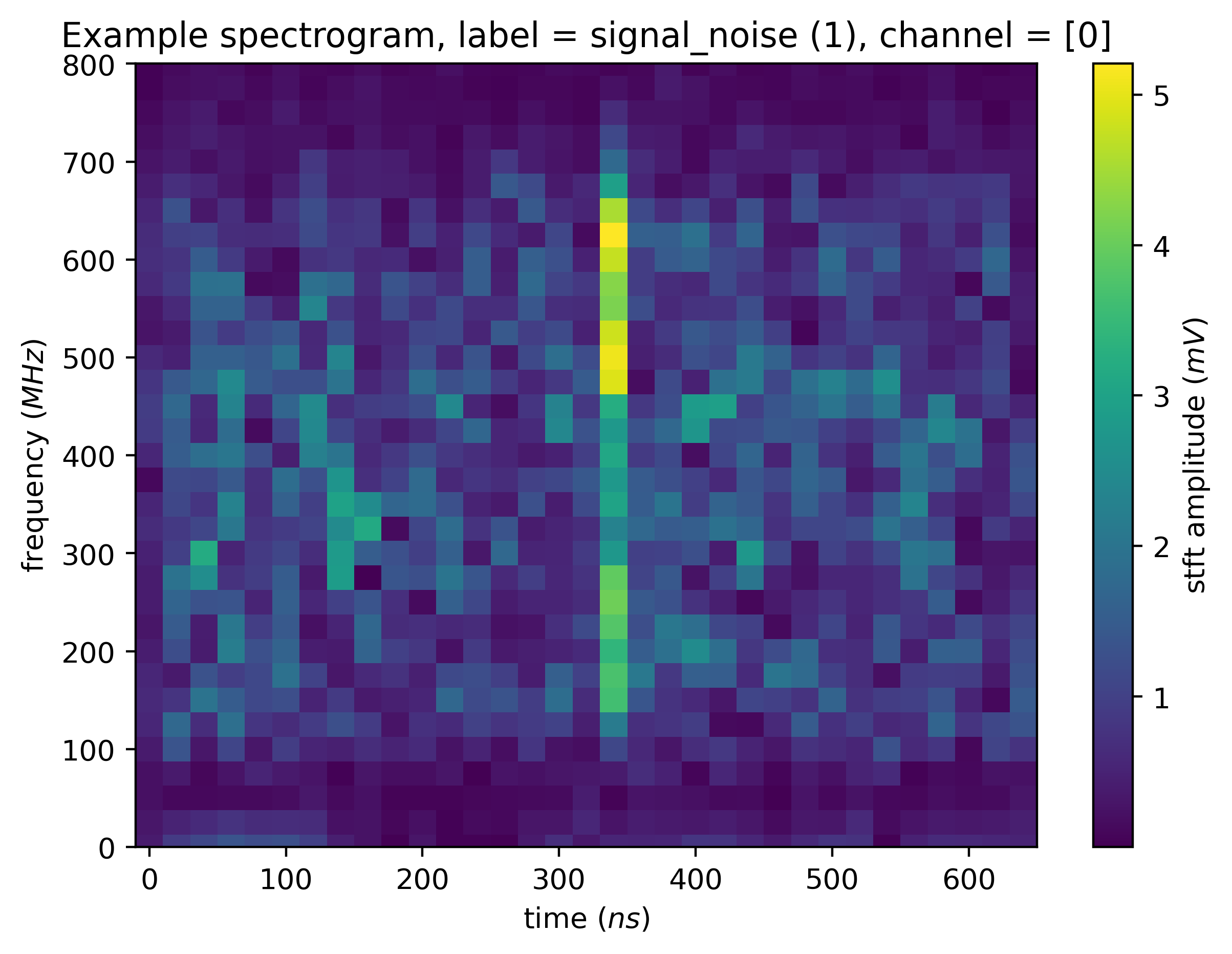}
    \caption{Example of a spectrogram representation of a noisy signal event. This is the direct result of applying a short-time Fourier transform on a single time trace. A full event would hence contain 24 of these spectrograms, one for each antenna at a single detector station.}
    \label{fig:spectrogram}
\end{figure}

The signal induced by a neutrino through the Askaryan effect exhibits a unique shape in both the time and frequency domain. To exploit the distinct features in both domains, the neural network is trained using a spectrogram-based data representation. An example of the signal class in this representation can be seen in Figure \ref{fig:spectrogram}.

One spectrogram represents data from one antenna. One data event contains data from all antenna and hence contains several spectrograms that are passed to the network at once. The first layers of the network architecture scan over each antenna individually. The final layers combine the features of all antenna by scanning over all spectrograms simultaneously, one for each antenna in an RNO-G station. This is especially important to make full use of the phased array antennas. The current architecture was trained only on the phased array antennas, since these make up the physical trigger, but the network was made flexible to easily include information from other antennas.

\begin{figure}[h!]
    \centering
    \includegraphics[width=0.9\linewidth]{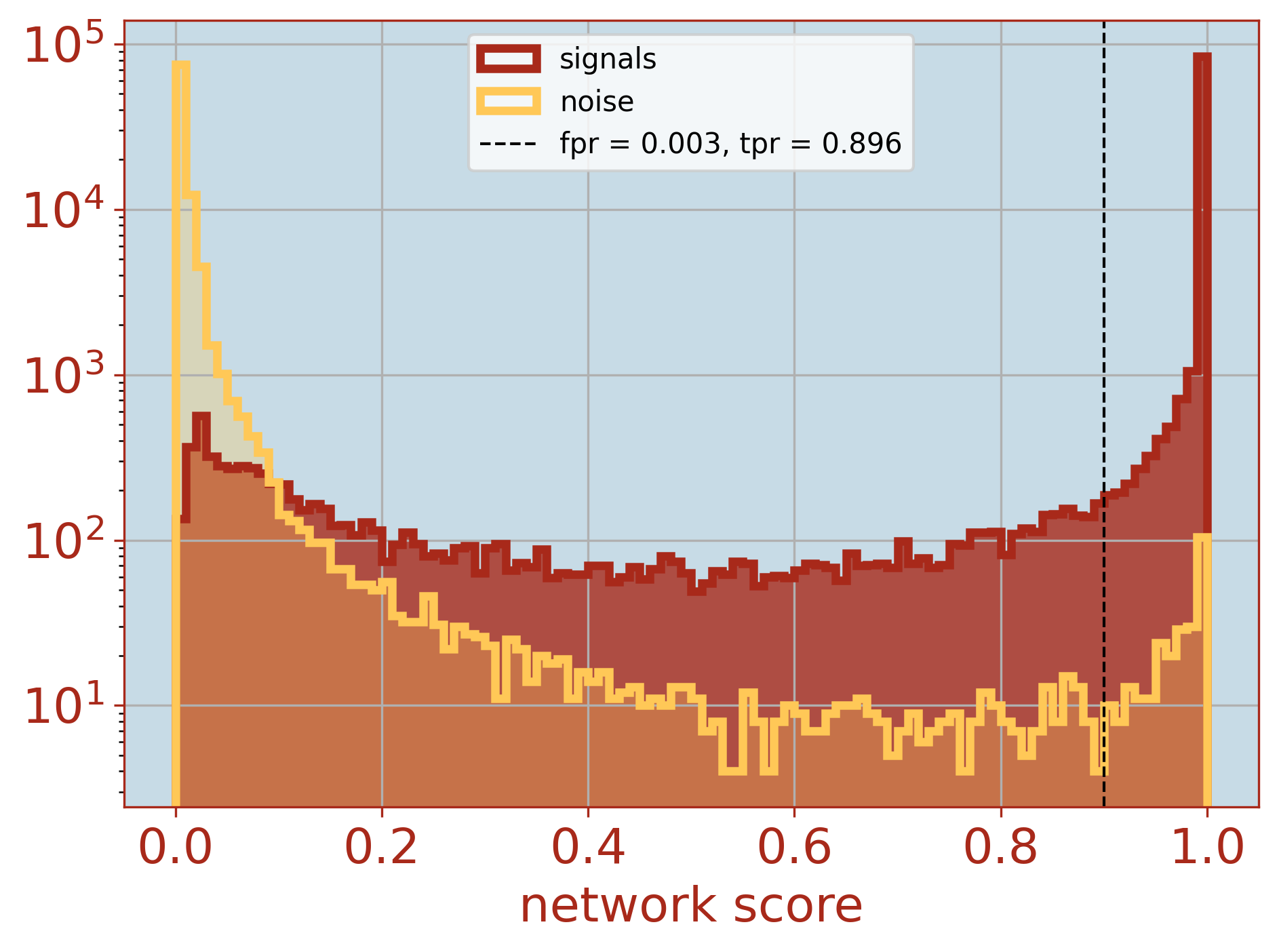}
    \caption{Evaluation of the event filter's performance by ways of comparing the score distributions given to events labeled as signals in red and events labeled as noise in yellow. The network score is a measure of the "belief" of the network that the given event is a signal. A perfect filter would score all noise events a $0$ and all signal events a $1$. One can choose a cutoff to convert this continuous scoring into a binary classification. This choice is free and can be adapted to obtain the filter's best performance. In this figure a cut-off of $0.9$ was chosen, which yields a false positive rate of $0.003$.}
    \label{fig:scores}
\end{figure}

To assess the network's performance in rejecting thermal noise, we construct a metric based on the score given by the network to an event and comparing that to the "true" label for a validation subset of the dataset. The distribution of the network's score can be seen in Figure \ref{fig:scores}.

As expected by design, the network exhibits superior performance in recognizing noise compared to signals. The network can be set to be arbitrarily stringent by choosing the minimum score required to label an event as a signal-like event. For example one can choose this in function of the required size of the resulting dataset. The false positive rate (fpr) of the network can be used to estimate the remaining noise population in the resulting dataset and is defined as the ratio of noise labeled as signal over the number of noise events. For a minimum score of $0.9$, this results in a false positive rate of $0.3 \%$.

\begin{figure}[h!]
    \centering
    \includegraphics[width=\linewidth]{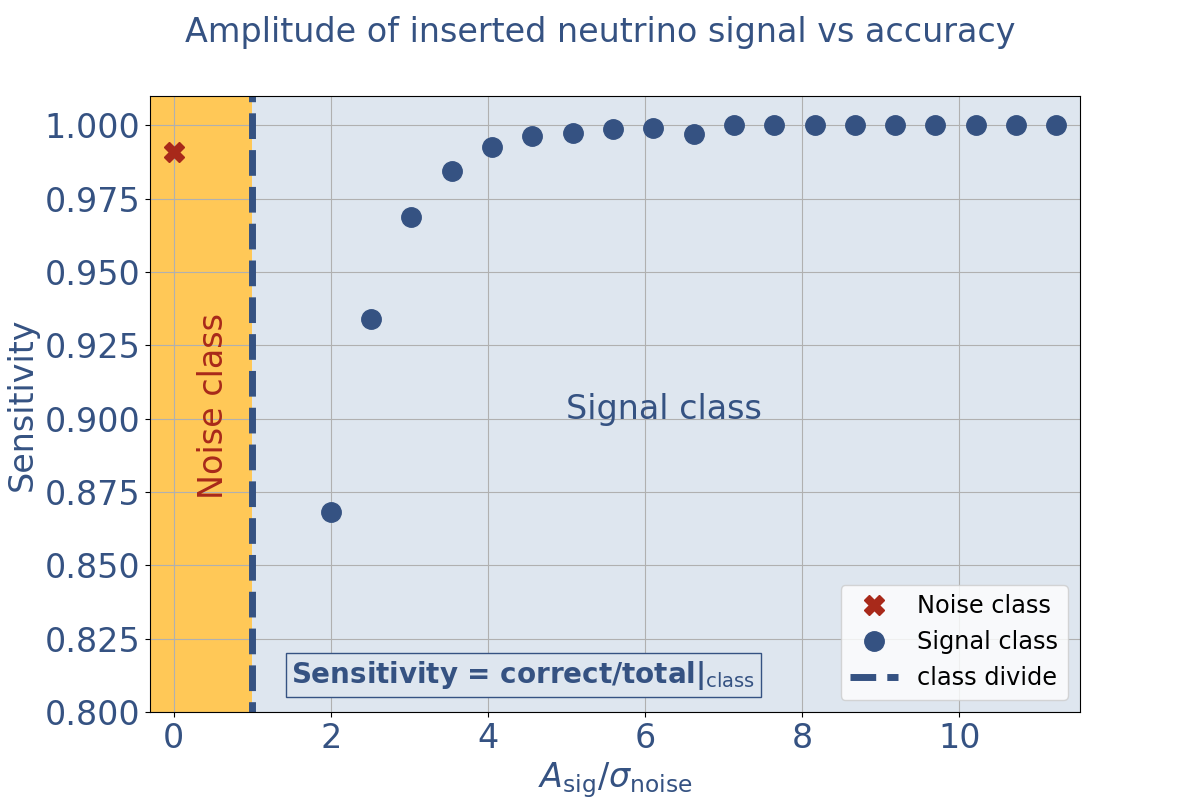}
    \caption{Sensitivity of the event filter in function of the signal to noise ratio of the given signal class. Sensitivity is designed as the number of correctly labeled events divided by the total number of given events. Note that a signal-to-noise ratio of $0$ corresponds to the noise class since these events, by definition, have no signal. To generate this plot, the criteria for an event to be labeled as signal was chosen to be: network score $> 0.5$.}
    \label{fig:accuracy}
\end{figure}

It is also possible to construct a metric better representing the influence of a signal's amplitude on the ability of the network to differentiate signal and noise. When constructing a signal event from data and simulation, the signal is randomly scaled to obtain a realistic signal-to-noise ratio, making the network less susceptible to overfitting. The metric can be seen in figure \ref{fig:accuracy}.

The sensitivity of the network (the ratio of correctly labeled events vs total events) is shown in function of signal-to-noise ratio. One can see that the noise class reaches high sensitivity, while lower signal amplitudes perform very poorly. The false positive rate on this plot can be interpreted as the distance between the noise class sensitivity, indicated by a red cross, and $1$. The goal is to minimize this distance while retaining a sufficiently high sensitivity for events in the signal class.

In conclusion, an offline filter was built based on a convolutional neural network architecture to categorize events measured at RNO-G as "noise" or "neutrino-like signal with noise". The focus of the filter is on rejecting noise events to generate a dataset for use in more sophisticated event reconstruction analyses. The current false positive rate, which can be interpreted as a metric for noise passing the filter, is $\sim 0.3 \%$. Efforts to improve the network's performance and lower this false positive rate are ongoing.

\section*{Acknowledgments} 
This research was performed under the umbrella of the RNO-G collaboration. Ruben Camphyn is a FRIA grantee of the Fonds de la Recherche Scientifique - FNRS.

\newpage

\bibliographystyle{apalike}
\bibliography{bibliography}

\end{document}